\documentstyle[12pt]{article}
\begin{document}
\topmargin 0pt
\oddsidemargin 7mm
\headheight 0pt
\topskip 0mm
\addtolength{\baselineskip}{0.40\baselineskip}

\hfill SNUTP 95-006

\hfill January, 1995

\hfill hep-th/9502115

\begin{center}
\vspace{0.5cm}
{\large \bf Hawking Radiation and Energy Conservation in an Evaporating
Black Hole}
\end{center}
\vspace{1cm}

\begin{center}
Won Tae Kim$^{(a)}$ and Julian Lee$^{(b)}$ \\
{\it Center for Theoretical Physics and Department of Physics, \\
Seoul National University, Seoul 151-742, Korea}
\end{center}

\vspace{1cm}

\begin{center}
{\bf ABSTRACT}
\end{center}

We define the Bondi energy for two-dimensional
dilatonic gravity theories by
generalizing the known expression of the ADM energy.
We show that our definition of the Bondi energy
is exactly the ADM energy minus the radiation energy at null infinity.
An explicit calculation is done for the evaporating black hole in the RST model
with the Strominger's ghost decoupling term. It is shown that the infalling
matter energy is completely recovered through the Hawking radiation and
the thunderpop.

\vspace{2cm}
\hrule
\vspace{0.5cm}
\hspace{-0.6cm}$^{(a)}$ E-mail address : wtkim@phyb.snu.ac.kr \\
\hspace{-0.6cm}$^{(b)}$ E-mail address : lee@phyb.snu.ac.kr \\
\vspace{1cm}
PACS number(s):04.60.+n, 97.60.Lf\\
\newpage

\pagestyle{plain}

\begin{center}
\vspace{0.5cm}
{\bf  I. INTRODUCTION}
\end{center}

 The two-dimensional
dilaton gravity
theories [1-5] are interesting as toy models for studying many interesting
issues in four-dimensional gravity theories. They possess most of the
interesting properties of the four-dimensional gravity theories
such as the existence of the black hole
solutions and Hawking radiations, and at the same time they are more amenable
to
quantum treatments than their four-dimensional counterparts.

  The evaporating black hole solution related to the
Hawking radiation [6] is particularly interesting since it is the situation
considered by Hawking in the context of the four-dimensional general
relativity when he put forward the famous information loss puzzle.
His radical proposal for the quantum mechanical evolution associated
with the black holes [7] has not yet been solved.
However, a recent work on the black-hole evaporation and back reaction
of the metric presented by Callan, Giddings, Harvey, and Strominger (CGHS)
[1], has shown that the quantum-mechanical gravity puzzles are no more
beyond our reach [8,9].

Russo, Susskind, and Thorlacius (RST)
obtained the RST model by adding
a local covariant counter term to the CGHS model [2,3]. The resulting
semiclassical equation is exactly solvable and describes the back reaction of
the metric at the one-loop level in the large $N$ limit where $N$ is a number
of conformal matter fields.
In particular, the model has an exact solution describing
the evaporation of black hole via Hawking radiations.
It has a mild violation of cosmic censorship hypothesis due to the naked
singularity as an isolated event [3].

On the other hand, it is a well known fact that one cannot construct (ordinary)
conserved
stress-energy-momentum tensor in general relativity except for space-times
having particular symmetries [10].
The fact that the stress-energy-momentum tensor for the matter fields alone is
not
conserved is not surprising since they exchange energies and momenta with
the gravitational field.
Furthermore, there is no notion corresponding
to the conserved stress-energy-momentum
tensor because a generally covariant tensor can only satisfy the covariant
conservation law.
However, one can introduce the concept of stress-energy-momentum for
gravity theories if we take the view that the general relativity
can be treated as a
spin-2 field theory in the Minkowski background [10,16].
Then, stress-energy-momentum will be a pseudotensor in the sense it is not
generally covariant but is
Lorentz covariant with respect to the Minkowski background metric.
Just as in the case of four-dimensional Einstein gravity, we can show that the
pseudotensor corresponding to energy density is a total derivative for the
two-dimensional dilaton gravity theories.
Therefore, for asymptotically flat space-times,
the energy becomes a surface term defined at either spatial or
null infinity. The former case is the Arnowitt-Deser-Misner (ADM) energy [11]
and the latter is just the Bondi energy [12].
Then, it is obvious that the difference of ADM and Bondi energy is
the integral of the current flowing out to null infinity.
In the four-dimensional Einstein gravity,
this current is interpreted as a radiation energy density.
In the case of two-dimensional dilaton gravity theories, the graviton and
dilaton fields have no propagating degrees of freedom and only the matter
radiation is capable of escaping to null infinity.

In Ref.\,[1], the Hawking radiation without the back reaction of the metric
can be calculated by using
the conformal anomaly of the
energy-momentum tensor of the matter part by imposing suitable boundary
conditions,
which is given by
\begin{equation}
<T^{\rm f}_{--} (\sigma^+, \sigma^-)>|_{\sigma^+ \rightarrow \infty} =
\frac{\lambda^2}{48} \left[ 1 - \frac{1}{(1+ \frac{m}{\lambda} e^{\lambda (
\sigma^- - \sigma^+_0 )})^2 } \right]
\end{equation}
where the metric in the tortoise coordinate $\sigma^\pm$ is
asymptotically Minkowskian at the future null infinity.
As was emphasized in Ref.\,[1], the total Hawking radiation is divergent. This
result it in contradiction to the energy conservation law, which is not
surprising since the effect of the back reaction to the geometry is not taken
into account.

Then, what about the energy conservation in the formation and
evaporation of the two-dimensional dilatonic black hole when we
consider the back reaction of the metric?
In this paper, we will consider the energy conservation in the two-dimensional
dilaton gravity theories, especially for the RST model.
In Sec. II, we define the notion
of the Bondi energy by generalizing the known expression
for the ADM energy for the dilatonic gravity models.
In Sec. III, we will calculate the Hawking radiation rate and the integrated
Hawking radiation for
the evaporating black hole in the RST model. In order to get the
positive definite Hawking flux [14], we include the Strominger's ghost
decoupling
term [15]. Then, we obtain a desirable result,
the total outgoing radiation due to the Hawking radiation and the classical
thunderpop being equal
to the energy of the infalling matter fields.
It means that the infalling
matter energy is completely recovered in the RST model.
In Sec. IV, we calculate the Bondi energy and the Hawking radiation
at null infinity, and
show that the total energy is conserved.
Finally, some discussions are given in Sec. V.

\begin{center}
\vspace{0.5cm}
{\bf  II. ADM, BONDI, AND RADIATION ENERGY IN TWO-DIMENSIONAL DILATON
GRAVITIES}
\end{center}

In this section, we present the definitions of the ADM and Bondi energy,
and their relation.
We consider dilaton gravity theories described by the action,
\begin{eqnarray}
S_{\rm T} &=& S_{\rm DG} + S_{\rm f}+ S_{\rm qt}, \label{start} \\
S_{\rm DG}&=& \frac{1}{2\pi} \int d^2 x \sqrt{-g} \left[e^{-2\phi} (R + 4 ({\bf
                \bigtriangledown} \phi)^2 +4\lambda^2) \right],  \\
S_{\rm f}&=& \frac{1}{2\pi} \int d^2 x \sqrt{-g} \left[ -\frac{1}{2}
\sum^N_{i=1}
                  (\bigtriangledown f_i)^2 \right], \\
S_{\rm qt}&=& \frac{1}{2\pi} \int d^2 x \sqrt{-g} \left[ -\frac{\kappa}{2}
             \phi R - \frac{Q^2}{2} R \frac{1}{\Box} R \right] \label{anomaly}
\end{eqnarray}
where $\kappa=Q^2=0$ gives the classical action, $\kappa=0$,
$2Q^2=\frac{N}{12}$
gives the CGHS model [1], and $\kappa=2Q^2=\frac{(N-24)}{12}$ gives
the RST model [2].
One then splits the above action as
\begin{equation}
S_{\rm T} = S_{\rm DG} + S_{\rm M}
\end{equation}
where $S_{\rm M}=S_{\rm f} +S_{\rm qt}$.
Then, the equation of motion for
the metric is given by
\begin{eqnarray}
G_{ \mu \nu}& =& T^{\rm M}_{\mu \nu}, \label{eom}\\
G_{ \mu \nu} &=&  \frac{2\pi}{\sqrt{-g}} \frac{ \delta S_{\rm DG} }{ \delta
g^{\mu \nu} } = 2e^{-2\phi}\left[ {\bf \bigtriangledown}_\mu  {\bf
\bigtriangledown}_\nu \phi + g_{\mu \nu} ( ({\bf \bigtriangledown} \phi )^2 -
\Box \phi -\lambda^2 ) \right], \label{einstein} \\
T^{\rm M}_{ \mu \nu }  &=& -\frac{2\pi} {\sqrt{-g}} \frac{ \delta S_{\rm M} }{
\delta g^{\mu \nu} }\label{stress},
\end{eqnarray}
where $T^{\rm M}_{\mu \nu}$ is the stress-energy-momentum tensor composed
of classical and quantum
matter parts.

In order to obtain the ordinary conserved quantity
instead of the covariant conserved one,
we expand the metric and dilaton fields around the linear dilaton vacuum (LDV),
\begin{eqnarray}
\label{eq:linear}
g_{\mu \nu}&=&\bar{g}_{\mu \nu} + h_{\mu \nu}, \quad \phi = \bar{\phi} + \psi,
\\
\bar{g}_{\mu \nu}&=& \eta_{\mu \nu},~~~~~~~~~~~
\bar{\phi} = -\lambda x^\alpha \eta_{\alpha \beta} \epsilon^\beta
\end{eqnarray}
where $\bar{g}_{\mu \nu}$ and $\bar{\phi}$ are the LDV configuration, and
$\eta_{11} = -\eta_{00}=1$ and $\epsilon^\alpha$ satisfies
$\epsilon^\alpha \eta_{ \alpha \beta} \epsilon^\beta = 1$.
One then linearizes the equation of motion (\ref{eom}) [16]:
\begin{equation}
G^{(1)}_{ \mu \nu} =  T^{\rm M}_{\mu \nu} - G^{(2)}_{\mu \nu}  \label{linear}
\end{equation}
where $G^{(1)}_{\mu \nu}$ is the linear part of $G_{\mu \nu}$ in $h_{\mu
\nu}$ and $\psi$ expansions, and $G^{(2)}_{\mu \nu}$ is the rest. If we take
the
time and
space coordinate $(t,q)$ such that  $x^\alpha \eta_{ \alpha \beta}
\epsilon^\beta
= q$, then it is straightforward to show that the left hand side of
(\ref{linear}) identically satisfies the conservation law,
\begin{equation}
\partial_\mu G^{(1) \mu 0} = 0, \label{lc}
\end{equation}
thanks to the linearized Bianchi identity [13].
Note that the total momentum, $\int dq G^{(1) 0 1}$,
is not conserved since the translational symmetry in the spatial
direction is spontaneously broken by the LDV.
Then, the right-hand side of (\ref{linear}) can be
thought of as the energy-momentum tensor for the fields,
$f$, $\phi$, and $g$.
Defining the matter current as
\begin{equation}
J^\mu =  T_{\rm M}^{\mu 0}  - G^{(2) \mu 0},
\end{equation}
we see that it satisfies the conservation law
\begin{equation}
\partial_\mu J^\mu =0 \label{lcc}
\end{equation}
due to Eq.\,(\ref{linear}) and Eq.\,(\ref{lc}).
If we consider a space-time which approaches the LDV
at spatial infinity fast enough, then
$J^1$ would vanish at $q \to \infty$ and the total energy,
\begin{eqnarray}
E_{\rm ADM}(t) &\equiv& \int_{-\infty}^\infty dq  J^{0}(t,q) \\
               &=& \int_{-\infty}^\infty dq  G^{(1) 0 0}(t,q), \label{adm}
\end{eqnarray}
is a conserved quantity, which is also called the ADM energy(mass).

After some straightforward algebra, the expression (\ref{adm}) reduces to
\begin{equation}
\label{old}
E_{\rm ADM} = 2 e^{2 \lambda q} (\partial_q \psi + \lambda \frac{h_{11}}{2}
) \vert_{q \to \infty}.
\end{equation}
 The contribution from $q \to  -\infty$ is easily seen to be zero for a
space-time which asymptotically approaches the LDV.
In the conformal gauge,
$g_{\mu \nu}=e^{2 \rho} \eta_{\mu \nu}$,
assuming the following asymptotic field configuration as $q \to \infty$,
\begin{equation}
\label{bcq}
\rho \approx A(t)e^{-2\lambda q},~~~~\psi \approx A(t)e^{-2 \lambda q},
\end{equation}
we obtain the expression
\begin{equation}
E_{\rm ADM} = 2 e^{2 \lambda q} (\partial_q  \psi + \lambda \rho )|_{q \to
\infty}
\label{confold}
\end{equation}
by keeping only the linear term in $\rho$. Note that the expression
(\ref{confold}) is valid also for any coordinate choice which approaches the
conformal gauge at infinities fast enough.
It is just the expression
used often in the literatures [17,18].

To define the Bondi energy, we need the boundary conditions at null infinity.
We require that
\begin{equation}
\label{bcpm}
\psi \approx D(y^-)e^{-\lambda y^+}, \quad \rho \approx D(y^-)e^{-\lambda y^+}
\end{equation}
where we used the light-cone coordinate, $y^{\pm}=t \pm q$.
For $y^+ \to -\infty$, it is enough to require that the configuration
approaches the LDV.
We now define the Bondi energy $B(y^-)$ as the energy evaluated along
the null line,
\begin{eqnarray}
\label{bondi}
B(y^-) &=& \frac{1}{2}\int_{-\infty}^\infty dy^+ G^{(1) - 0} (y^+, y^- )
      \nonumber \\
&=& 2e^{\lambda (y^+ - y^-)} ( \partial_+ - \partial_- +
\lambda)  \psi  \vert_{y^+ \to \infty}.
\end{eqnarray}
Note that the Bondi energy is defined at the null infinity while
the ADM energy is defined at the spatial infinity.

Now we show the difference between
$E_{\rm ADM}$ and $B(y^-)$. Obviously, it can be
represented by
the integral of $G^{(1) + 0}$ along the null line,
from the point
$(\infty,~-\infty)$  to the point $(\infty,~y^-)$,
\begin{eqnarray}
\label{diff}
E_{\rm ADM}-B(y^-) &=& - \frac{1}{2}\int_{-\infty}^{y^-} dy^-  G^{(1)+0} (y^+,
y^-) |_{y^+ \to \infty}
\nonumber \\
                  & =& - \int_{-\infty}^{y^-}
dy^- \partial_-\left( 2e^{ \lambda (y^+ - y^-) } ( \partial_+ -
\partial_- +
\lambda)  \psi  \right) \vert_{y^+ \to \infty}
\end{eqnarray}
which is just an identity.
After some calculation, the right-hand side of
(\ref{diff}) can be identified by the
integral of the radiation flux
of matters $T^{\rm M}_{--}$ at null infinity by using (\ref{linear}) under the
boundary condition (\ref{bcpm}) [13], $i.e.$,
\begin{equation}
\label{relation}
E_{\rm ADM} -B(y^-) =\int_{-\infty}^{y^-} dy^-
\left( T^{\rm f}_{--} + T^{\rm qt}_{--} \right)\vert_{y^+ \to \infty}.
\end{equation}
It is plausible to regard the quantum matter part of the radiation
as a Hawking radiation which is explicitly given by
\begin{eqnarray}
\label{hradiation}
h(y^-)& =& T^{\rm qt}_{--}\vert_{y^+ \to \infty}  \nonumber \\
      & =&-2Q^2\left( (\partial_- \rho)^2   -\partial_-^2 \rho +t_- (y^-)
\right)|_{y^+ \to \infty}
          + \frac{\kappa}{4}(4 \partial_- \rho \partial_- \phi -2 \partial_-^2
\phi)
            |_{y^+ \to \infty} \nonumber \\
      &\approx & -2Q^2 t_-(y^-),
\end{eqnarray}
where the function $t_-(y^-)$ reflects the non-locality of
the conformal anomaly term of Eq.\,(\ref{anomaly}).
As a result, the radiation is composed of the classical (conformal)
matter and the Hawking radiation, and the energy conservation relation
(\ref{relation}) is
written as
\begin{equation}
\label{bt}
E_{\rm ADM}(t)-B(y^-) =\frac{1}{2} \sum_{i=1}^{i=N} \int_{-\infty}^{y^-}dy^-
(\partial_-f_i)^2|_{y^+ \to \infty}  -2Q^2 \int_{-\infty}^{y^-} dy^- t_-(y^-).
\end{equation}
Note that the Bondi energy is just the remaining energy after the classical and
quantum Hawking radiation has been emitted from the system.
We will explicitly study this energy conservation relation in the RST model.

\begin{center}
\vspace{0.5cm}
{\bf  III. HAWKING RADITION IN THE RST MODEL}
\end{center}

In this section, we apply the formal concepts developed in the above section
to the RST model by explicit calculations of relevant quantities.
 From the action (\ref{start}), the RST model with the Strominger's ghost
decoupling term $S_{\rm St}$ [15] in the conformal gauge is given by
\begin{eqnarray}
S&=&S_{\rm T} +S_{\rm St}\label{rstact}, \\
S_{\rm T}&=&\frac{1}{\pi} \int d^2 x \left[
e^{-2\phi} (2\partial_+ \partial_-
\rho -  4\partial_+ \phi \partial_- \phi +  \lambda^2 e^\rho)  \right.
\nonumber
\\
&+& \left. \frac{1}{2}
\sum^N_{i=1} \partial_+ f_i \partial_- f_i
-\kappa \partial_+ \rho \partial_-
\rho   -\kappa \phi \partial_+ \partial_-
\rho \right], \\
S_{\rm St}\hspace{-0.3cm} &=&\hspace{-0.3cm}  \frac{1}{\pi} \int d^2 x \left[ 2
\partial_+ (\rho -\phi)
\partial_- (\rho -\phi) \right].
\end{eqnarray}
We introduced $S_{\rm St}$ to make the Hawking radiation positive definite
for arbitrary $N$.
In the conformal gauge, one must impose two constraint equations corresponding
to the vanishing metric components:
\begin{eqnarray}
\label{const}
T_{\pm \pm}\hspace{-0.3cm} & =&\hspace{-0.3cm} (e^{-2 \phi} +\frac{\kappa}{4})(
4 \partial_\pm \rho \partial_\pm
\phi - 2 \partial^2_\pm \phi ) + \frac{1}{2} \sum^{N}_{i=1} \partial_\pm f_i
\partial_\pm f_i  - \kappa ( \partial_\pm \rho \partial_\pm \rho -
\partial^2_\pm
\rho)   \\    \nonumber
&+& 2 ( \partial_\pm(\rho -\phi) \partial_\pm (\rho -\phi) -\partial^2_\pm
(\rho -\phi) ) - \kappa t_\pm (x^\pm) =0
\end{eqnarray}
where the functions $t_\pm (x^\pm)$ are needed to satisfy asymptotic
physical boundary conditions.
Following the Bilal and Callan [4] and de Alwis's method [5], we perform
field redefinition to a Liouville theory [14],
\begin{eqnarray}
\Omega &=&\frac{\kappa}{2 \sqrt{\kappa -2}}
\phi +\frac{e^{-2\phi}}{\sqrt{\kappa -2}},  \nonumber    \\
\chi &=& \sqrt{\kappa -2} \rho -\frac{(\kappa -4)}{2\sqrt{\kappa -2}} \phi +
\frac{e^{-2\phi}}{\sqrt{\kappa -2}}.
\end{eqnarray}
Then, the action (\ref{rstact}) and the two constraints (\ref{const})
in terms of the redefined fields
are given by
\begin{eqnarray}
S &=&\frac{1}{\pi} \int d^2 x  \left[ -\partial_+ \chi \partial_- \chi
+ \partial_+ \Omega \partial_- \Omega + \lambda^2 e^{\frac{2}{\sqrt{\kappa
-2}}(\chi -\Omega)}
  + \frac{1}{2} \sum^N_{i=1} \partial_+ f_i \partial_- f_i \right], \label{RST}
\\
\kappa t_\pm &=& -\partial_\pm \chi \partial_\pm \chi + \partial_\pm \Omega
\partial_\pm \Omega
+\sqrt{\kappa -2} \partial^2_\pm \chi + \frac{1}{2} \sum^N_{i=1}
\partial_\pm f_i \partial_\pm f_i. \label{newconst}
\end{eqnarray}
 From the constraint equations (\ref{newconst}), we can determine the total
central
charge,
\begin{eqnarray}
c&=&c_\chi +c_\Omega +c_M +  c_{ghost}   \nonumber \\
 &=& [1-12(\kappa -2)] +1 +N -26,
\end{eqnarray}
together with the ghost contribution.
To impose the constraints consistently at the quantum level,
we fix $\kappa =\frac{N}{12}$
which is positive, and the Hawking radiation becomes positive definite.
 From the action (\ref{RST}), we obtain the equations of motion
\begin{eqnarray}
\partial_+ \partial_-\chi + \frac{\lambda^2}{\sqrt{\kappa -2}}
e^{\frac{2}{\sqrt{\kappa -2}} (\chi -\Omega)}&=& 0,    \\
-\partial_+ \partial_-\Omega - \frac{\lambda^2}{\sqrt{\kappa -2}}
e^{\frac{2}{\sqrt{\kappa -2}} (\chi -\Omega)}&=& 0,     \\
\partial_+ \partial_- f_i &=&0.
\end{eqnarray}
In the Kruskal gauge, $\rho =\phi$, the static solution is given by
\begin{eqnarray}
\Omega(x^+, x^-)& =&\chi(x^+, x^-) \nonumber \\
                & =&-\frac{\lambda^2}{\sqrt{\kappa -2}} x^+ x^-
                    +P\frac{\kappa}{\sqrt{\kappa -2}}\ln(-\lambda^2 x^+x^-)
                    +\frac{m}{\lambda\sqrt{\kappa -2}}
\end{eqnarray}
where $P$ and $m$ parametrize different solutions.
For $P=-\frac{1}{4}$ and $m=0$, it becomes LDV solution,
and for $P=0$ and $m \neq 0$, it is a thermal equilibrium solution
since the metric is independent of time in appropriate coordinates.

Let us now consider an evaporating black-hole solution formed by an incoming
shock
wave at $x^+ = x^+_0$ given by $T^{\rm f}_{++}=\frac{m}{\lambda x^+_0}
\delta (x^+ -x^+_0)$ [2],
\begin{eqnarray}
\Omega(x^+, x^-)\hspace{-0.3cm}& =&\hspace{-0.3cm}\chi(x^+, x^-) \nonumber \\
              \hspace{-0.3cm}  &
=&\hspace{-0.3cm}-\frac{\lambda^2}{\sqrt{\kappa
-2}} x^+ x^-  \nonumber\\
   &-&\frac{1}{4} \frac{\kappa}{\sqrt{\kappa -2}}\ln(-\lambda^2
x^+x^-)
                    -\frac{m}{\lambda x_0^+\sqrt{\kappa -2}}(x^+ - x^+_0)
                    \Theta(x^+ -x^+_0) \label{sol}
\end{eqnarray}
where the matching condition at $x^+=x^+_0$ are obtained by the ++
constraint equation (\ref{newconst}) with the above incoming pulse wave.
Then, the singularity can form at $\phi_c =-\frac{1}{2}\ln\frac{\kappa}{4}$
where $\frac{d\Omega(\phi_c)}{d\phi}=0$. The singularity occurs at the
boundary of the range of $\Omega$ where $\Omega(\phi_c)=
\frac{\kappa}{4\sqrt{\kappa -2}} [1- \ln \frac{\kappa}{4}]$. From
(\ref{sol}), the
curve $\phi(\bar{x}^+, \bar{x}^-) = \phi_c $ is given by
\begin{eqnarray}
1-\ln\frac{\kappa}{4}&\hspace{-0.4cm} =
\hspace{-0.4cm}&-\frac{4\lambda^2}{\kappa} \bar{x}^+\bar{x}^- -\ln(-\lambda^2
\bar{x}^+ \bar{x}^-)    \nonumber \\
\hspace{-0.4cm}&-&\hspace{-0.4cm}\frac{4m}{\lambda x^+_0 \kappa}
(\bar{x}^+ - x^+_0)\Theta(\bar{x}^+ - x^+_0).\label{curve}
\end{eqnarray}
This is the same form as in the case without the Strominger term except for the
change
of the value of $\kappa$.
The location of the singularity is inside an apparent horizon which is given by
$\partial_+
\phi =0$. The apparent horizon gives another curve:
\begin{equation}
\label{another}
\hat{x}^+(\hat{x}^- +\frac{m}{\lambda^3 x^+_0} )
+\frac{\kappa}{4\lambda^2}=0.
\end{equation}

Following the suggestion of Hawking [19], RST showed that the singularity
and apparent horizon collide in a finite proper
time and the singularity
is naked after the two have merged [2].
 From (\ref{curve}) and (\ref{another}), the intersection point is given by
\begin{eqnarray}
x^+_s & = &\frac{\kappa \lambda x^+_0}{4m} ( e^{\frac{4m}{\kappa \lambda}} -1),
\nonumber \\
x^-_s & = & -\frac{m}{\lambda^3 x^+_0} \frac{1}{(1-e^{-\frac{4m}{\kappa
\lambda}})}\>.
\end{eqnarray}
As shown by RST, it is possible to match the evaporating solution (\ref{sol})
with
a shifted LDV solution at the null line, $x^-=x^-_s$.

The conformal transformation, $x^\pm = \pm \frac{1}{\lambda}e^{\pm \lambda
\sigma^\pm}$, does not give an asymptotically static configuration
and in particular the
dilaton and graviton fields do not approach the correct form of LDV at
infinity,
so we introduce
a quasi-static coordinate $y^\pm$ where the fields approach LDV in spatial and
null
infinities in both $q \to \pm \infty$ and $y^+ \to \pm \infty$ [4],
\begin{equation}
x^+=\frac{1}{\lambda} e^{\lambda y^+},~~~x^-=-\frac{1}{\lambda}
e^{-\lambda y^-} -\frac{m}{\lambda^3 x^+_0}\Theta(y^+ - y^+_0) \label{quasi}.
\end{equation}
In this coordinate, $\Omega$ and $\chi$ are static to the leading order. Note
that this coordinate transformation is not conformal due to the presence of the
$\Theta$ function, however, it does not matter since
the coordinate transformation asymptotically goes to conformal gauge at
infinities.
We denote the intersection point in this coordinate by $(y^+_s, y^-_s)$,
\begin{equation}
y^+_s =\frac{1}{\lambda}\ln(\lambda x^+_s),~~~y^-_s=-\frac{1}{\lambda}
\ln(-\lambda x^-_s  -\frac{m}{\lambda^2 x^+_0} )
\end{equation}
which will be used in later.
The Penrose diagram of the RST model is depicted in Fig. 1.

Let us now consider the Hawking radiation. From
the fundamental condition that $T_{\pm\pm}$ must be a true tensor
without anomaly, we require the anomalous transformation as
\begin{equation}
t_\pm (y^\pm) = \left( \frac{\partial y^\pm}{\partial \sigma^\pm}\right)^{-2}
\left( t_\pm (\sigma^\pm) - \frac{1}{2} D^s_{\sigma^\pm} (y^\pm) \right)
\end{equation}
where $D^s_{\sigma^\pm}(y^\pm)$ is the Schwarzian derivative.
Then, following [4], we obtain the Hawking radiation,
\begin{eqnarray}
h(y^-)&=&-\kappa t_-(y^-) \nonumber \\
&=&\frac{\kappa \lambda^2}{4}
\left[ 1- \frac{1}{(1+\frac{m}{\lambda} e^{\lambda(y^- -y^+_0)})^2} \right]
\label{h}
\end{eqnarray}
for $y^- < y^-_s$ and $h(y^-) =0$ for $y^- > y^-_s$.
A typical form of the Hawking radiation is illustrated in Fig. 2.
Note that the expression for the Hawking radiation (\ref{h}) is same
as (\ref{hradiation}) by identifying $2Q^2=\kappa$.
As expected, for $y^- \to -\infty$, there is no Hawking radiation.
In the limit $ y^- \rightarrow y^-_s -0$, the radiation is
\begin{equation}
h(y^-_s-0) =\frac{\kappa \lambda^2}{4} (1-e^{-\frac{8m}{\kappa \lambda}}).
\end{equation}

For $y^- < y^-_s$, the integrated Hawking flux $H(y^-)$, is calculated
as
\begin{eqnarray}
H(y^-)&=&\int^{y^-}_{-\infty} dy^- h(y^-) \nonumber \\
      &=&\frac{\kappa \lambda}{4}\left[ 1-\frac{1}{ (1+\frac{m}{\lambda}
      e^{\lambda(y^- -y^+_0)}) } +\ln(1+\frac{m}{\lambda}
      e^{\lambda(y^- -y^+_0)})  \right] \label{hr}.
\end{eqnarray}
For the interesting limit, $y^- \rightarrow y^-_s-0$, we obtain
\begin{equation}
\label{hrs}
H(y^-_s-0)=m +\frac{\kappa \lambda}{4}(1 -e^{-\frac{4m}{\kappa \lambda} } )
\end{equation}
which is greater than the total energy of the infalling matter field.
This point is clarified in the next section.

On the othe hand, for $y^- >y^-_s$, $H(y^-)$ is given by
\begin{eqnarray}
H(y^-)&=& \int_{-\infty}^{y^-_s -0} dy^- h(y^-) +\int_{y^-_s +0}^{y^-}dy^-
             h(y^-)    \nonumber \\
      &=& H(y^-_s -0) +0  \nonumber \\
      &=& H(y^-_s +0)   \label{epsi}.
\end{eqnarray}
Therefore, the integrated Hawking flux $H(y^-)$ is saturated when the
black hole is completely evaporated and
the total Hawking flux is a just $H(y^-_s).$

\begin{center}
\vspace{0.5cm}
{\bf IV. ENERGY CONSERVATION IN THE RST MODEL}
\end{center}

In this section, we calculate the Bondi energy, and
prove the energy conservation at the arbitrary time in the RST model.

Let us first show the solution (\ref{sol}) satisfies the boundary
conditions (\ref{bcq}) and (\ref{bcpm}) in the asymptotically
quasi-static coordinates (\ref{quasi}).
For $y^+ > y^+_0$, the evaporating black-hole solution (\ref{sol})
can be written as the following form,
\begin{eqnarray}
\label{soly}
\Omega(y^+,y^-)&=&\frac{1}{\sqrt{\kappa -2}} ( e^{\lambda (y^+ -y^-)}
+\frac{m}{\lambda} e^{\lambda (y^+ -y^+_0)} )
                -\frac{\kappa}{4\sqrt{\kappa -2}} \ln( e^{\lambda (y^+ -y^-)}
+\frac{m}{\lambda} e^{\lambda (y^+ -y^+_0)} ) \nonumber \\
               &-&\frac{m}{\lambda \sqrt{\kappa -2}}(e^{\lambda (y^+
-y^+_0)}-1),\\
\label{solyx}
\chi(y^+,y^-)&=&\Omega(y^+,y^-) + \frac{\lambda \sqrt{\kappa -2}}{2} (y^+-y^-),
\end{eqnarray}
and the vacuum solution $(\bar{\Omega},\bar{\chi})$ is
\begin{eqnarray}
\bar{\Omega}(y^+,y^-)&=&\frac{1}{\sqrt{\kappa -2}}e^{\lambda(y^+ -y^-)}
-\frac{\kappa \lambda}{4\sqrt{\kappa -2}} (y^+-y^+_0),  \label{vacsol} \\
\bar{\chi}(y^+,y^-)&=& \bar{\Omega}(y^+,y^-) +\frac{\lambda \sqrt{\kappa
-2}}{2} (y^+ -y^-).
\end{eqnarray}
The asymptotic behaviors of the solution (\ref{soly}) at the spatial and null
infinity are,
\begin{eqnarray}
\left( e^{-2\phi} +\frac{\kappa}{2} \phi \right) \vert_{q \to
\infty}&\hspace{-0.3cm}= \hspace{-0.3cm}&
e^{2 \lambda q} -\frac{\kappa \lambda}{2} q  + \frac{m}{\lambda} +
O(e^{-\lambda q}), \label{q}  \\
\left( e^{-2\phi} +\frac{\kappa}{2} \phi \right) \vert_{y^+ \to
\infty}&\hspace{-0.3cm}=\hspace{-0.3cm}&
e^{\lambda(y^+ -y^-)}\!-\!\frac{\kappa \lambda}{4}(y^+ -y^-)\hspace{-0.1cm}
-\hspace{-0.1cm}\frac{\kappa}{4}
\ln \left[ \left(1+\frac{m}{\lambda} e^{\lambda(y^- - y^+_0)}\right)
\hspace{-0.1cm}
e^{-\frac{4m}{\kappa \lambda}} \right]. \label{yplus}
\end{eqnarray}
 From (\ref{q}) and (\ref{yplus}), the asymptotic forms $\phi$ are obtained:
\begin{eqnarray}
\phi \vert_{q \to \infty}&=&(\bar{\phi} + \psi) \vert_{q \to \infty} =
-\lambda q + A(t) e^{-2 \lambda q} + \cdots,            \\
\phi \vert_{y^+ \to \infty}&=&(\bar{\phi} + \psi) \vert_{y^+ \to \infty}=
-\frac{\lambda}{2} (y^+ -y^-) +D(y^-) e^{- \lambda y^-} +\cdots
\end{eqnarray}
where $A(t)=-\frac{m}{2\lambda}$ and $D(y^-)=\frac{\kappa}{8}
e^{\lambda y^-} \ln  \left[ \left( 1+\frac{m}{\lambda} e^{\lambda(y^- - y^+_0)}
\right)
e^{-\frac{4m}{\kappa \lambda}} \right]$.
Therefore the solution (\ref{sol}) satisfies the boundary conditions
(\ref{bcq})
and
(\ref{bcpm}) in the quasi-static coordinate.
Then, it is convenient to write the Bondi energy as
\begin{equation}
B(y^-) = \sqrt{\kappa -2}(\lambda +\partial_- -\partial_+)\delta \Omega
(y^+,y^-)|_{y^+ \rightarrow +\infty}. \label{newbondi}
\end{equation}
For solution satisfying the boundary condition (\ref{bcpm}), it is
straightforward to
show that the Bondi energy (\ref{newbondi}) reduces to the previous form
(\ref{bondi}) since
\begin{eqnarray}
&&\sqrt{\kappa -2}(\lambda +\partial_- -\partial_+)\delta \Omega
(y^+,y^-)|_{y^+ \rightarrow +\infty} \nonumber \\
&=& (\lambda +\partial_- -\partial_+)
\left[ \frac{\kappa}{2} \psi + e^{\lambda (y_+ - y_-)}(e^{-2\psi}-1)
\right]_{y^+ \rightarrow +\infty} \nonumber \\
&=& (\lambda +\partial_- -\partial_+ )\left[ e^{\lambda (y_+ - y_-) }(-2\psi+2
\psi^2)\right]_{y^+
\rightarrow +\infty} \nonumber \\
&=& 2 e^{\lambda (y_+ - y_-)}(\partial_+ - \partial_- + \lambda)(\psi
-\psi^2)\vert_{y^+
\rightarrow +\infty} \nonumber \\
&=& 2 e^{\lambda (y_+ - y_-)}(\partial_+ - \partial_- + \lambda)\psi
\vert_{y^+
\rightarrow +\infty}
\end{eqnarray}
where $\delta \Omega = \Omega - \bar{\Omega}$.
Similarly, the ADM energy can be
written as
\begin{equation}
\label{newadm}
E_{\rm ADM}(t) = \sqrt{\kappa -2}(\lambda -\partial_q)\delta \Omega |_{q
\rightarrow +\infty}.
\end{equation}

By putting Eq.\,(\ref{soly}) and Eq.\,(\ref{vacsol}) into (\ref{newadm}) and
(\ref{newbondi}),
we obtain
ADM energy and Bondi energy respectively,
\begin{eqnarray}
E_{\rm ADM}(t)&=&m,  \\
\label{bondie}
B(y^-)&=& m -\frac{\kappa \lambda}{4} \left[ \frac{ \frac{m}{\lambda}}{(
\frac{m}{\lambda} + e^{-\lambda(y^- -y^+_0)})}
+\ln(1 +\frac{m}{\lambda} e^{\lambda(y^--y^+_0)} )\right].
\end{eqnarray}
Note that at the point $y^-_s-0$, the Bondi energy is given by
\begin{equation}
\label{bondis}
B(y^-_s-0)=-\frac{\kappa \lambda}{4}(1-e^{-\frac{4m}{\kappa \lambda}})
\end{equation}
which is negative.
The behavior of the negative Bondi energy is shown in
Fig. 3 as an illustration.

We easily see that for $y^-< y^-_s$, the total (ADM) energy of the evaporating
black hole system is the sum of the integrated Hawking flux (\ref{hr}) and the
Bondi energy (\ref{bondie}):
\begin{eqnarray}
E_{\rm ADM}&=&B(y^-) +H(y^-)  \nonumber \\
        &=&m.  \label{total}
\end{eqnarray}
In this region, there is no classical contribution to the radiation.

For $y^- > y^-_s$,
we note that the classical negative energy thunderpop should be taken
into consideration. Indeed, integrating the energy density carried out by the
the thunderpop, we obtain [2]
\begin{eqnarray}
E_{\rm thunderpop} &=&\int^{y^-}_{-\infty} dy^- T^{\rm f}_{--} (y^-) \nonumber
\\
                   &=& -\frac{\kappa \lambda}{4}
                   (1-e^{-\frac{4m}{\kappa \lambda}}),
\end{eqnarray}
while for $y^- <y^-_s$, there is no thunderpop contribution.

There is no Bondi energy for $y^- > y^-_s$ because of $\delta \Omega = 0$,
( {\it i.e.}, we now
have
the LDV and the Bondi energy of the LDV is zero) and we have
\begin{eqnarray}
E_{\rm ADM}(t)&=&B(y^-) + H(y^-) + E_{\rm thunderpop} \nonumber \\
              &=& 0+ H(y^-_s +0) + E_{\rm thunderpop} \nonumber \\
              &=& m
\end{eqnarray}
by using $H(y^-)$ in (\ref{epsi}).
Therefore we see that for arbitrary $y^-$ the energy conservation relation
is valid:
\begin{equation}
E_{\rm ADM}(t) = B(y^-) + H(y^-) +\int^{y^-}_{-\infty} dy^-
T^{\rm f}_{--} (y^-).
\end{equation}
Therefore, we proved that the total energy of the infalling classical shock
wave
is preserved throughout the formation and subsequent evaporation of the
black hole.

\begin{center}
\vspace{0.5cm}
{\bf V. DISCUSSIONS}
\end{center}

In this paper, we expressed the Bondi energy which is consistent with the
usual definition of the Bondi energy as being the energy left in the system
after
the radiation has been occurred [10]. Another attempt to construct the
Bondi energy
was done by Bilal [20]. However, we are puzzled by the fact it violates the
energy
conservation. He applied his definition of the Bondi energy to the example of
evaporating black hole in the RST model, which we also considered.
His Bondi energy is
positive definite up to the point where the thunderpop is emitted. However, if
the Bondi energy
is defined as the energy left in the system after the radiation,
then it should be negative just before the emission of the negative
thunderpop energy in the RST model.
On the other hand, our Bondi energy is not
necessarily positive definite. We believe our definition of the Bondi energy
is more
reasonable since it satisfies the usual requirements for the Bondi energy and
the energy conservation. Also, the change of Bondi energy could not be followed
exactly up to the end point of the Hawking radiation and the
calculation was done only to the leading order in $m$ in Ref.\,[20].
In this paper, we could do
the calculations exactly up to the endpoint.

Finally, we comment on the Hawking radiation $h(y^-_s-0)
=\frac{\kappa \lambda^2}{4} [1-e^{-\frac{8m}{\kappa \lambda}}]$ just before the
end point of the black hole evaporation.
It is definitely positive due to the decoupling of ghost contribution and
depends on the total energy $m$. For $m \rightarrow \infty$, it recovers
the well-known two-dimensional result without back reaction of the metric
,${\it i.e.}$, $h(y^-_s-0) \rightarrow \frac{\kappa
\lambda^2}{4}$. This result is natural in that the
infinitely large mass of matter fields effectively generates the
 static solution
since the large mass of black hole may radiate eternally as far as $m$ is
infinite.
Another limit one can consider is
for $\kappa \lambda >> m$, where one may $properly$ neglect
the subleading quantum effect in $\frac{1}{N}$ expansions among
the one-loop graphs.
In this case, the Hawking radiation depends on the linear power of mass,
$h(y^-_s-0) \rightarrow 2\lambda m$.

\section*{ACKNOWLEDGEMENTS}
We are very grateful to Choonkyu Lee for helpful discussions.
We were supported by the Korea Science
and Engineering Foundation through the Center for Theoretical Physics (1995).

\newpage
\section*{FIGURE CAPTIONS}
\begin{description}
\item[FIG. 1] Penrose diagram of the evaporating black hole in the RST model.
              An incoming shock wave at $y^+ =y^+_0$ produces the black
              hole and the negative energy thunderpop goes out at
                             $y^-=y^-_s$. The zigzag line denotes the
singularity.
\item[FIG. 2] A plot of the Hawking radiation ${h(y^-) \over \kappa \lambda^2}$
               up to $y^-_s \simeq 4$
              for the case ${m \over \lambda}=\lambda x^+_0=1$. $h(y^-)=0$ for
                       $y^->y^-_s$.
\item[FIG. 3] A plot of the Bondi energy ${B(y^-) \over \kappa \lambda}$ up to
                $y^-_s \simeq 4$
              for the case ${m \over \lambda}=\lambda x^+_0=1$.
              $B(y^-)=0$ for $y^->y^-_s$.
\end{description}

\newpage
\end{document}